\documentclass[journal]{IEEEtran}

\ifCLASSINFOpdf
\else
   \usepackage[dvips]{graphicx}
\fi
\usepackage{url}

\usepackage{graphicx}
\usepackage{amsmath}
\usepackage{algorithm}
\usepackage{algorithmic}

\begin{document}

\title{Fuzzy SLIC: Fuzzy Simple Linear Iterative Clustering}

\author{Chong Wu$^1$, \IEEEmembership{Student Member, IEEE}, Jiangbin Zheng$^2$, Zhenan Feng$^2$, Houwang Zhang$^2$, Le Zhang$^3$, Jiawang Cao$^3$ and Hong Yan$^4$, \IEEEmembership{Fellow, IEEE}
\thanks{This work was supported by the Hong Kong Research Grants Council
[Project C1007-15G] and City University of Hong Kong [Projects 7005230 and 9610460].}
\thanks{Chong Wu and Hong Yan are with the Department of Electrical Engineering, City University of Hong Kong, Kowloon, Hong Kong (e-mail: chongwu2-c@my.cityu.edu.hk \& h.yan@cityu.edu.hk).}
\thanks{Jiangbin Zheng is with the School of Infomatics, Xiamen University, Xiamen, 361005, China (e-mail: jiangbinzheng@stu.xmu.edu.cn).}
\thanks{Zhenan Feng, Houwang Zhang and Jiawang Cao are with the School of Automation, China University of Geosciences, Wuhan 430074, China  (e-mail: fengzhenan@cug.edu.cn\& zhanghw@cug.edu.cn \& CJW@cug.edu.cn).}
\thanks{Le Zhang is with the Department of Computer Science and Technology, Tongji University, Shanghai 200092, China (e-mail: zhangle\_tj@163.com).}
\thanks{$^1$The first author designed and implemented the algorithm, designed the experiment and the ablation study, run most experiments, and wrote the paper. $^2$The second to the forth authors contributed equally in main experiment, data analysis, and ablation study to this work. $^3$The fifth to the sixth authors contributed equally in main experiment. $^4$The seventh author gave some constructive comments and revised the paper.}
\thanks{Copyright \copyright 2020 IEEE. Personal use of this material is permitted. However, permission to use this material for any other purposes must be obtained from the IEEE by sending an email to pubs-permissions@ieee.org.}

}

\markboth{This paper has been accepted by IEEE Transactions on Circuits and Systems for Video Technology. DOI: 10.1109/TCSVT.2020.3019109}
{Shell \MakeLowercase{\textit{et al.}}: Bare Demo of IEEEtran.cls for IEEE Journals}
\maketitle

\begin{abstract}
Most superpixel methods are sensitive to noise and cannot control the superpixel number precisely. To solve these problems, in this paper, we propose a robust superpixel method called fuzzy simple linear iterative clustering (Fuzzy SLIC), which adopts a local spatial fuzzy C-means clustering and dynamic fuzzy superpixels. We develop a fast and precise superpixel number control algorithm called onion peeling (OP) algorithm. Fuzzy SLIC is insensitive to most types of noise, including Gaussian, salt and pepper, and multiplicative noise. The OP algorithm can control the superpixel number accurately without reducing much computational efficiency. In the validation experiments, we tested the Fuzzy SLIC and OP algorithm and compared them with state-of-the-art methods on the BSD500 and Pascal VOC2007 benchmarks. The experiment results show that our methods outperform state-of-the-art techniques in both noise-free and noisy environments.
\end{abstract}


\IEEEpeerreviewmaketitle

\section{Introduction}
\noindent The concept of superpixel was first introduced in \cite{Ren2003}. Superpixel methods group pixels similar in color and other properties \cite{SLIC}. They capture the redundancy, abstract and preserve the structure from the image \cite{SLIC,JFast,FuzzyS}. Through substituting thousands of pixels to hundreds of superpixels, they also improve the computational efficiency of subsequent image processing tasks \cite{SLIC,Harmony,SEEDS}. With the above advantages, superpixel methods are widely used as a preprocessing technique in many image processing tasks like template matching \cite{H8}, image quality assessment \cite{W8}, image classification \cite{FuzzyS,Z6}, image segmentation \cite{FuzzyS, luc,A7,R7,kumar}, \emph{etc.} 

\begin{figure*}[h]
\centering 
\centerline{\includegraphics[width=1\linewidth]{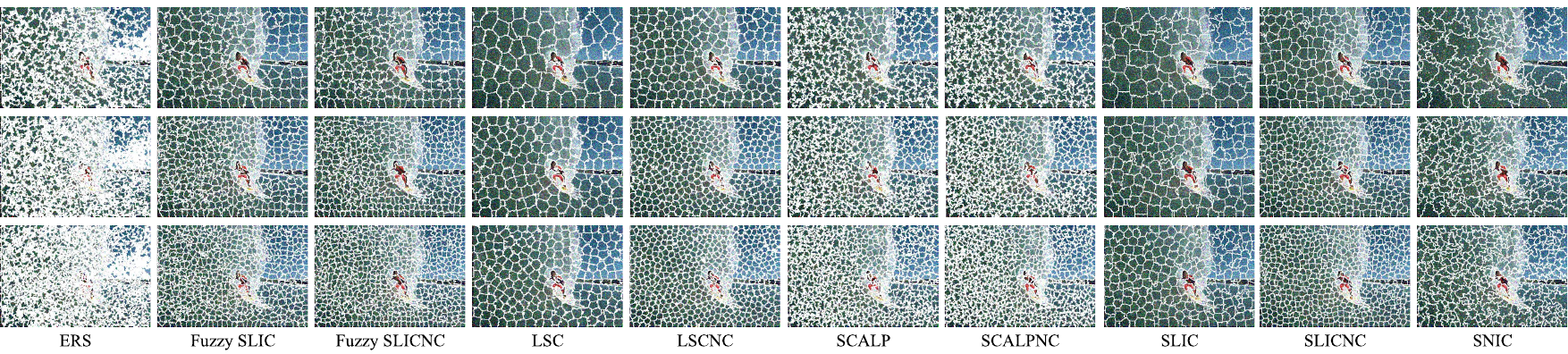}}
\caption{Visual comparison results containing salt and pepper noise with noise density 0.2. Methods with `NC' in their names are the ones using OP algorithm to control the superpixel number and all methods use the same $m = 200$ (upper), $m = 400$ (middle), and $m=600$ (bottom). Fuzzy SLIC and Fuzzy SLICNC outperform other methods in terms of regularity and boundary accuracy while another robust superpixel method SCALP performs even worse than LSC, LSCNC, and SLICNC. In its superpixel segmentation, the boundaries are rougher than the boundaries obtained by compared methods. It means that it cannot handle salt and pepper noise. There is an interesting phenomenon that all methods after using OP algorithm can not only generate precise superpixel number but also become more robust than original ones.}
\label{fig19}
\end{figure*}

\begin{figure*}[h]
\centering 
\centerline{\includegraphics[width=1\linewidth]{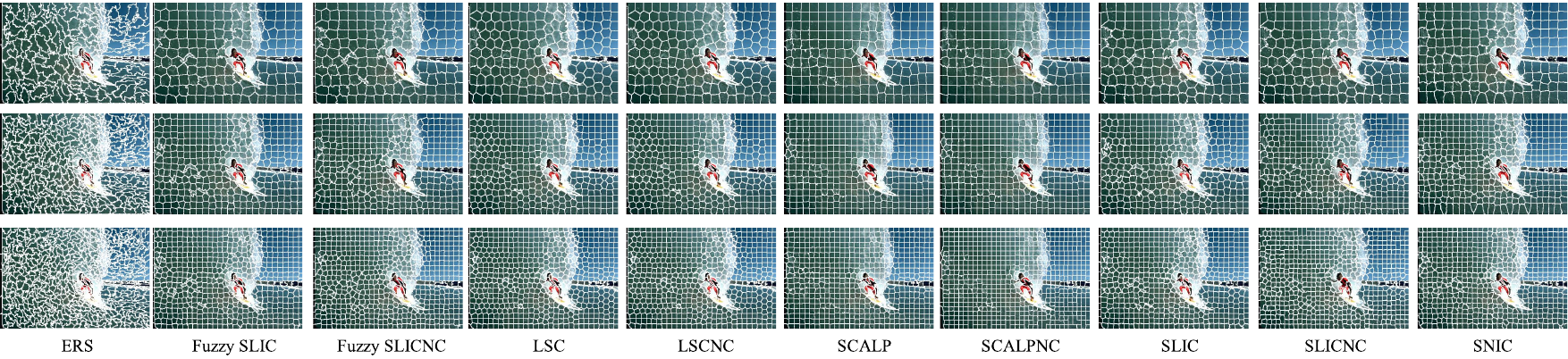}}
\caption{Visual comparison results in noise-free environment. Methods with `NC' in their names are the ones using OP algorithm to control the superpixel number and all methods use the same $m = 200$ (upper), $m = 400$ (middle), and $m=600$ (bottom). All methods except ERS achieve similar performance in terms of boundary accuracy and regularity.}
\label{fig191}
\end{figure*}

\begin{figure*}[h]
\centering 
\centerline{\includegraphics[width=1\linewidth]{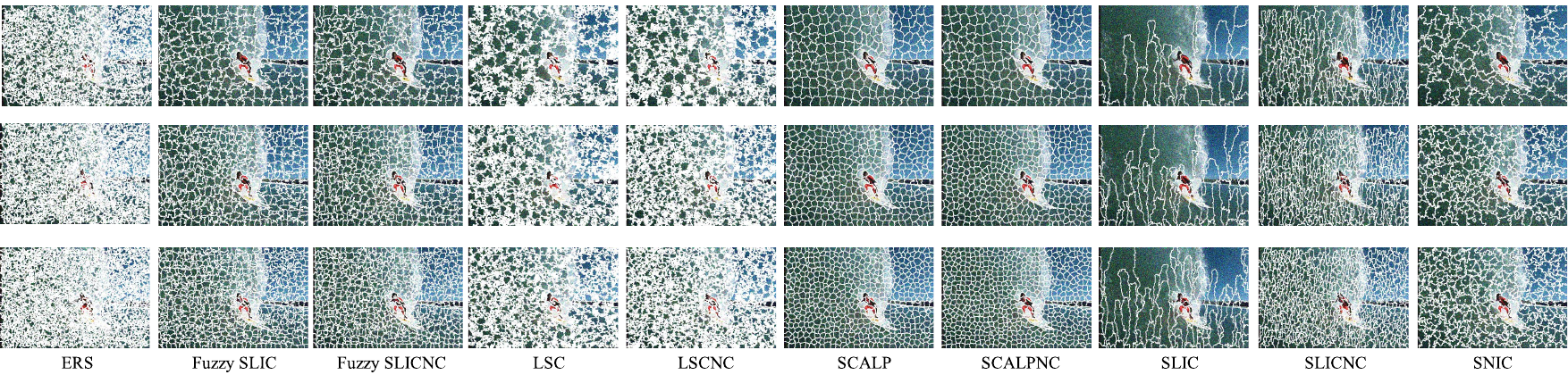}}
\caption{Visual comparison results containing Gaussian noise with noise level 0.2. Methods with `NC' in their names are the ones using OP algorithm to control the superpixel number and all methods use the same $m = 200$ (upper), $m = 400$ (middle), and $m=600$ (bottom). SCALP and SCALPNC achieve the best performance with this noise. Fuzzy SLIC and Fuzzy SLICNC achieve the second place. The boundary accuracy of Fuzzy SLIC and Fuzzy SLICNC are still better than most methods and very close to SCALP and SCALPNC. As to the regularity of superpixel, the performance of Fuzzy SLIC and Fuzzy SLICNC is also only inferior to SCALP and SCALPNC and better than other comparison methods.}
\label{fig192}
\end{figure*}

\begin{figure*}[h]
\centering 
\centerline{\includegraphics[width=1\linewidth]{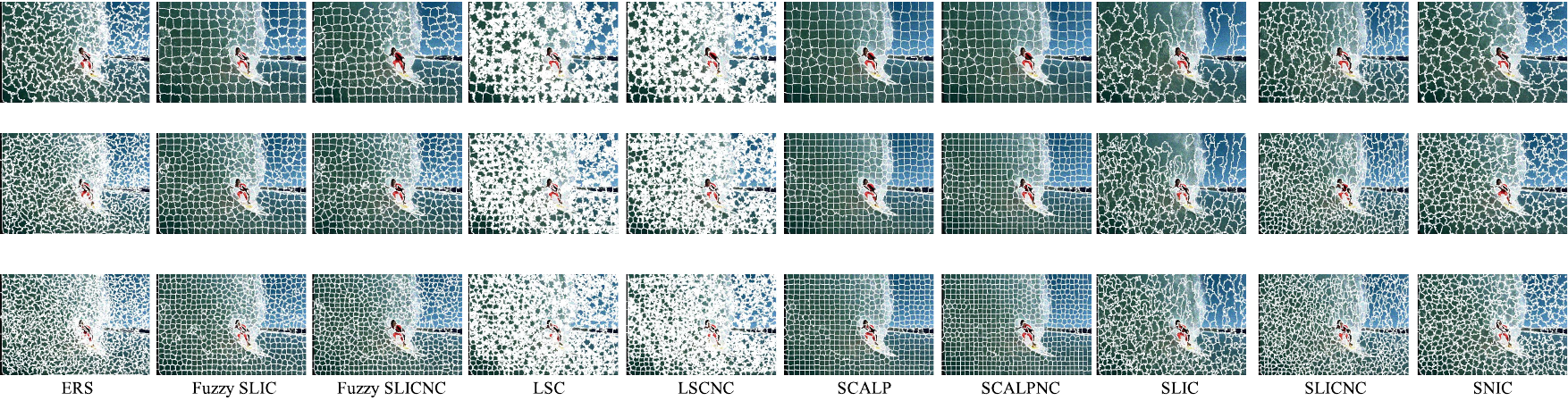}}
\caption{Visual comparison results containing multiplicative noise with noise level 0.2. Methods with `NC' in their names are the ones using OP algorithm to control the superpixel number and all methods use the same $m = 200$ (upper), $m = 400$ (middle), and $m=600$ (bottom). Fuzzy SLIC, SCALP, and SCALPNC achieve the best performance with this noise, while Fuzzy SLICNC achieves the second place. The difference between the first rank and the second rank is not significant.}
\label{fig193}
\end{figure*}

\begin{figure*}[h]
\centering 
\centerline{\includegraphics[width=1\linewidth]{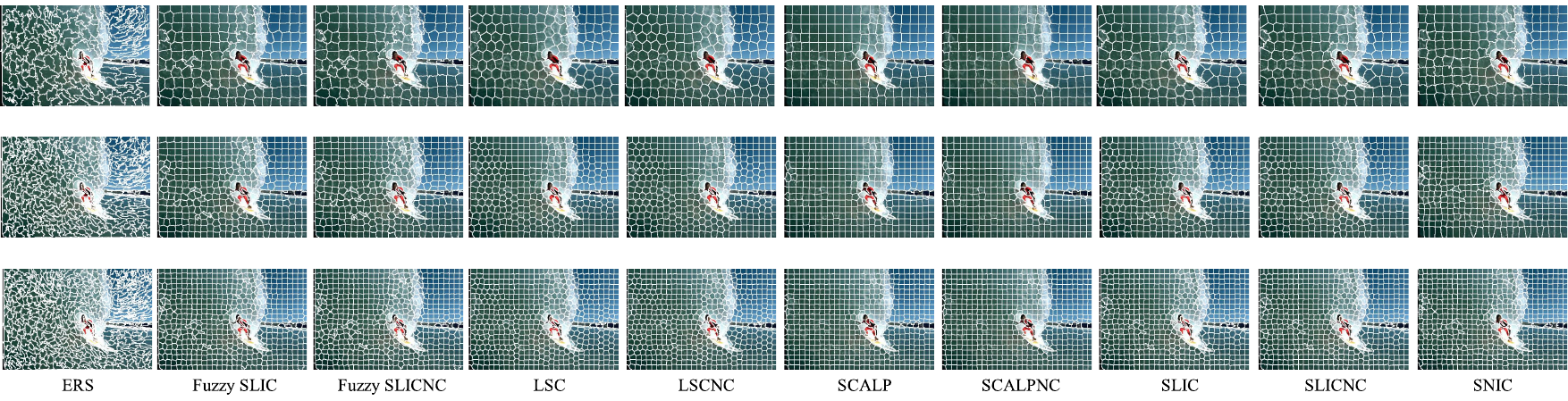}}
\caption{Visual comparison results after using a simple mean filter. Methods with `NC' in their names are the ones using OP algorithm to control the superpixel number and all methods use the same $m = 200$ (upper), $m = 400$ (middle), and $m=600$ (bottom). All methods achieve similar results to their performance in noise-free environment.}
\label{fig194}
\end{figure*}

A good superpixel method should have properties, like compactness, partition, connectivity, boundary adherence, low computational complexity, high memory efficiency, controllable superpixel number, and precise generated superpixel number \cite{SLIC}. To achieve these properties, various efficient superpixel methods have been proposed. They can be broadly classified into two categories: graph-based and clustering-based methods \cite{SSN}. Graph-based methods consider the original image as an undirected graph and use edge-weights obtained as the similarity of color to partition this graph \cite{SUB}. Normalized cuts (NC) \cite{Ren2003} and entropy rate superpixels (ERS) \cite{ERS} are two typical graph-based superpixel methods. NC adopts cuts while ERS exhibits a bottom-up merging of pixels to generate superpixels \cite{SUB}. Most of graph-based methods are unable to control the compactness of superpixels and suffered from high computational complexity \cite{SUB}. Clustering-based methods are often used as a preprocessing method for image segmentation tasks \cite{FuzzyS,luc,kumar}. They consider superpixel generation as a clustering problem. They take color, spatial distance, depth, \emph{etc}. as the clustering features. Simple linear iterative clustering (SLIC) \cite{SLIC}, linear spectral clustering (LSC) \cite{LSC}, simple noniterative clustering (SNIC) \cite{SNIC}, superpixels with contour adherence using linear path (SCALP) \cite{SCALP}, and superpixel sampling networks (SSN) \cite{SSN} are typical clustering-based methods. Compared to graph-based methods, clustering-based ones have the advantages of faster speed and controllable compactness. However, most of them need an additional post-processing step to enforce the connectivity \cite{SUB,SNIC}. It makes them unable to generate precise number of superpixels as required \cite{SNIC}. Moreover, when applied as preprocessing techniques, superpixel methods often have substantially degraded performance with noisy images \cite{SCALP} as Figs. \ref{fig19}-\ref{fig194} show. Recently, several advanced superpixel methods have been proposed, such as bilateral geodesic distance (BGD) \cite{BGD}, adaptive nonlocal random walks (ANRW) \cite{ANRW}, and superpixel optimization using higher order energy (SOHOE) \cite{HOE}, but they are not designed for noisy images and their performance for noisy images still need significant improvement. Although some methods \cite{SCALP,WAT} have been proposed to resist noise, they are still not robust enough. Some of them are just designed for specific types of noise only and some others just use a preprocessing step of denoising.


To solve the robustness problem, in this paper, we introduce a local spatial fuzzy C-means clustering method and dynamic fuzzy superpixels, which is called fuzzy simple linear iterative clustering (Fuzzy SLIC). Compared to state-of-the-art methods, Fuzzy SLIC does not need a preprocessing step of denoise and still has a favorable speed. In addition, we introduce a fast and precise superpixel number control algorithm called the onion peeling (OP) algorithm for clustering-based superpixel methods. To validate the proposed methods, we tested Fuzzy SLIC and OP algorithm on Pascal VOC2007 \cite{voc2007} and BSD500 \cite{BSD500} and compared them with ERS, LSC, SCALP, SLIC, and SNIC in five noise conditions: (1) non-noise; (2) salt and pepper noise; (3) Gaussian noise; (4) multiplicative noise; (5) mean filter. The results show that our methods outperform state-of-the-art methods in both noisy and noise-free environment. The contributions of this paper include
\begin{itemize}
  \item [1.] 
  This is the first attempt of introducing fuzzy C-means clustering to the clustering-based superpixel method. 
  \item [2.]
  We have investigated dynamic fuzzy superpixels to efficiently improve the computational performance of the proposed method.
  \item [3.]
  We have developed a robust superpixel method which has high performance against different types of noise. 
  \item [4.]
  A fast and accurate superpixel number control algorithm is proposed for clustering-based superpixel methods. 
  \end{itemize}


\section{Fuzzy Simple Linear Iterative Clustering and Onion Peeling Algorithm}
In this section, we will introduce two main parts of the proposed Fuzzy SLIC and a fast and precise superpixel number control algorithm.

\subsection{Local Spatial Fuzzy C-means Clustering}
Fuzzy C-means clustering (FCM) \cite{FCM} is widely used in the field of image segmentation \cite{kumar,SFCM,FRFCM}.  However, standard FCM is sensitive to some independent noise points \cite{SFCM}. To overcome this problem, some researchers proposed spatial constrained fuzzy C-means clustering methods (SFCMs) \cite{SFCM,SKFCM}. Through introducing spatial information to FCM, SFCMs can yield homogeneous regions and are more robust against noise than the standard FCM \cite{SFCM}. In this paper, we propose a local spatial fuzzy C-means clustering (LSFCM) which uses a local search instead of a global search. Different to FCM, because of using a local search, the objective function of LSFCM is as follow,
\begin{equation}
\label{e1}
\min J(\boldsymbol{U},\boldsymbol{c}) = \sum_{j=1}^{z}\sum_{i=1}^{N_j}u^{t}_{ij}\|x_{i}-c_{j}\|^{2},
\end{equation}
\begin{equation}
\label{e12}
\boldsymbol{U} \in M_{fc},
\end{equation}
\begin{equation}
\label{ex12}
\begin{split}
M_{fc}=\{&\boldsymbol{U}\in V_{zN_j}|u_{ij}\in[0,1] \ \forall i,j;\\
&\sum_{j=1}^{z}u_{ij}=1\ \forall i;\\
&0<\sum_{i=1}^{N_j}u_{ij}<N_j\ \forall j\},
\end{split}
\end{equation}
where, $\boldsymbol U$ is the fuzzy partition matrix integrated with spatial information which is different to the original FCM, $N_j$ is the number of pixels which are in a grid searching region of cluster $j$, $c_j$ is the centroid of cluster $j$, $z$ is the number of clusters after initialization which is often different to input cluster number $m$, $t$ is the fuzzy partition matrix exponent which should be larger than 1 for controlling the degree of fuzzy overlap, $u_{ij}$ is the degree of membership of data point $i$ in the cluster $j$, and $\| \cdot \|$ denotes the inner product norm.

Change above optimization problem to an unconstrained optimization problem by introducing Lagrange multipliers as follow,
\begin{equation}
\label{e13}
F(\lambda,u_{ij}) = J(\boldsymbol{U},\boldsymbol{c}) + \lambda (\sum_{j=1}^{z}u_{ij}-1).
\end{equation}

%
%
%

By taking partial derivative of $F$ with respect to $u_{ij}$ and $\lambda$ in Eq. (\ref{e13}), we have
\begin{equation}
\label{e2}
u_{ij} = {\left (\sum_{k=1}^{m_i} \left(\frac{\|x_{i}-c_{j}\|}{\|x_{i}-c_{k}\|}\right)^{\frac{2}{t-1}}\right )}^{-1},
\end{equation}
where, $m_i$ is the number of possible labels of pixel $i$, because of the local search, only the pixel in a cluster's searching region is used to update $\emph{\textbf{u}}$. Details of the derivation are provided in the supplementary material.


In an image, neighboring pixels are usually highly correlated which means that the probability that these pixels belong to the same cluster is high \cite{SFCM}. Hence, introducing neighboring spatial information can improve the robustness of the clustering methods. LSFCM introduces the spatial information in its fuzzy partition matrix $\emph{\textbf{U}}$ to compute the new fuzzy partition matrix $\emph{\textbf{U'}}$. LSFCM adopts a $3 \times 3$ window centered at pixel $i$ to select its neighbors and uses the neighbor's fuzzy partition degree to construct the spatial function as follow,
\begin{equation}
\label{e4}
h_{ij}=\sum_{k\in \mathcal{W}_i}u_{kj},
\end{equation}
where, $\mathcal{W}_i$ is the set of the pixels selected by a square window which centers at pixel $i$. Then combine the $\emph{\textbf{H}}$ with $\emph{\textbf{U}}$ to obtain $\emph{\textbf{U'}}$, 
\begin{equation}
\label{e5}
u'_{ij} = \frac{u_{ij}^{p}h_{ij}^{q}}{\sum_{k=1}^{m_i}u_{ik}^{p}h_{ik}^{q}},
\end{equation}
where, $p$ and $q$ are the control parameters. A larger $q$ means more important of the effect from neighbors. In this paper, we adopt $p = 0$, $q = 2$ to achieve high robustness. The selection process of hyperparameters $p$ and $q$ is shown in Section \ref{Exp}. Through introducing spatial information to the fuzzy partition matrix, the original membership will be strengthened in a homogenous region. While for a noisy pixel, the weight of a noisy cluster will be reduced by the neighboring spatial information.

\begin{figure*}
\centering 
\centerline{\includegraphics[width=\linewidth]{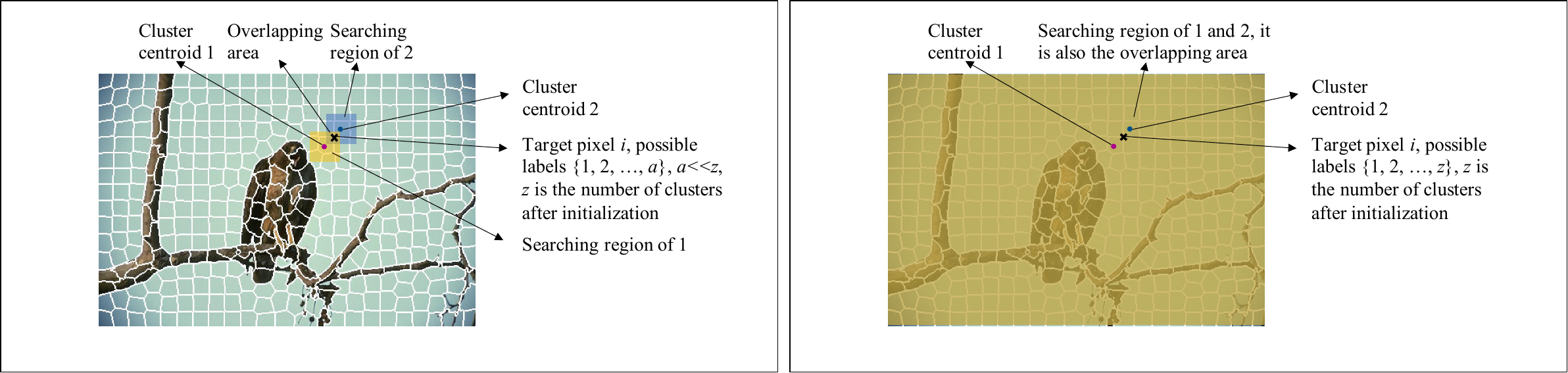}}
\caption{Comparison of local spatial fuzzy C-means clustering and SFCM \cite{SFCM}. The local here means the membership of a pixel is only shared by several nearest superpixels which is different to the SFCM \cite{SFCM}. The latter one uses a global search: the update of each centroid needs the information from all clusters which causes redundant information and loses the computational efficiency.}
\label{fig0}
\end{figure*}

Taking the partial derivative of $F$ with respect to $\boldsymbol{c}$ in Eq. (\ref{e13}) and replacing $\emph{\textbf{U}}$ with $\emph{\textbf{U'}}$, then we can update $\boldsymbol{c}$ as follow,
\begin{equation}
\label{e6}
c_{j}=\frac{\sum_{i=1}^{N_j} {u'}_{ij}^{t}x_{i}}{\sum_{i=1}^{N_j} {u'}_{ij}^{t}}.
\end{equation}
$\boldsymbol c$ is a weighted linear combination of $\boldsymbol x$, this is why our algorithm is called fuzzy simple linear iterative clustering. The demonstration of LSFCM is as Fig. \ref{fig0} shows.

\subsection{Dynamic Fuzzy Superpixels} 
\begin{figure}
\centering 
\centerline{\includegraphics[width=0.8\linewidth]{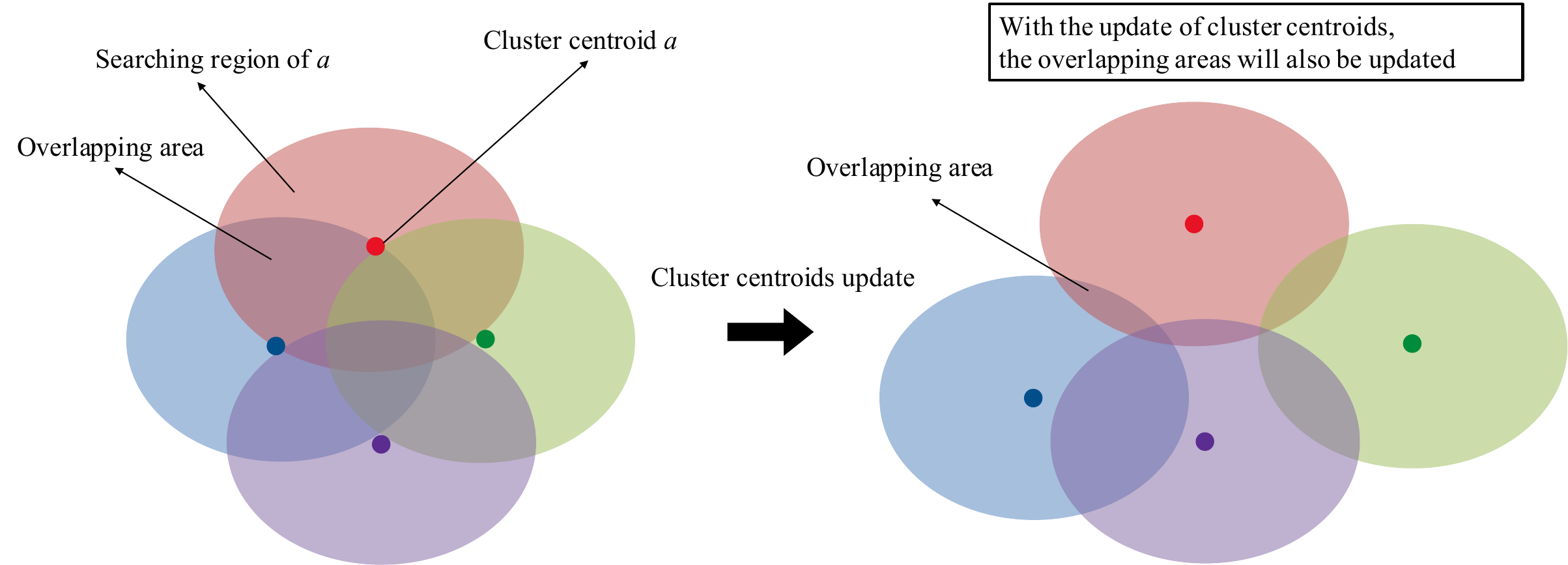}}
\caption{The searching region of each superpixel has some overlapping areas with other superpixels. Fuzzy SLIC performs a dynamic region evolution from start to end. The dynamic overlapping areas in clustering process are called dynamic fuzzy superpixels.}
\label{fig1x}
\end{figure}

How to define the overlapping searching region is a crucial step of LSFCM. We consider two ways: (1) static overlapping region (simple, fast, but less reasonable and not suitable for a clustering method), which is used in a study of improving the performance of TurboPixels \cite{Turbo} and is called fuzzy superpixel (FS) \cite{FuzzyS}; (2) dynamic overlapping region (reasonable but high computational and memory cost because each pixel has different number of possible labels as Fig. \ref{fig1x} shows and in the clustering process, the maximum number of possible labels of each pixel will also be changed in different iterations). In this paper, we consider dynamic overlapping region but modify it to reduce the computational and memory requirement. We call the dynamic overlapping part of each superpixel's searching region dynamic fuzzy superpixel (DFS). We find that most pixels do not have more than 3 labels in DFS. We tested different maximum number of possible labels, and we also find when the maximum number of labels of each pixel is set to 3, LSFCM will achieve the best balance of performance and computational cost against FS. The details of the process to determine the maximum number of labels of each pixel in DFS are shown in Section \ref{Exp}. With the introduction of DFS, the computational performance of LSFCM will be further improved without performance lost, and the Eq. (\ref{e6}) will be modified as follow,
\begin{equation}
\label{e8}
c_{j}=\frac{\sum_{i=1}^{\mathcal{N}_j} {u'}_{ij}^{t}x_{i}}{\sum_{i=1}^{\mathcal{N}_j} {u'}_{ij}^{t}},
\end{equation}
where, $\mathcal{N}_j$ is the number of pixels which have the label of $j$, this time, it is not a certain value, and different clusters or superpixels $\mathcal{S}$ will have different $\mathcal{N_S}$.

The pseudo-code of Fuzzy SLIC is shown in Algorithm \ref{alg4}. 

\begin{algorithm}[htb]
\scriptsize
\caption{Fuzzy simple linear iterative clustering}
\label{alg4}
\begin{algorithmic}
\REQUIRE 
Image matrix $\emph{\textbf{I}}$, number of clusters $m$, compactness coefficient $C$\\
\ENSURE 
Label matrix $\emph{\textbf{L}}$, number of superpixels $m_p$\\
\STATE 
Get the grid map with each grid size $\sqrt\frac{N}{m}$ of $\emph{\textbf{I}}$\\
\STATE 
Initialize the cluster centroids $\emph{\textbf{c}}$ by sampling pixels in each grid and update cluster number $z$\\
\STATE
Adjust $c_j$ to the lowest gradient position in its 8 spatial nearest neighbors\\
\STATE
$\emph{\textbf{L}}$[1:$N$] $\gets -1$, $\emph{\textbf{G}}$[1:$N$, 1:3] $\gets -1$, $\emph{\textbf{D}}$[1:$N$, 1:3] $\gets \infty$ ($\emph{\textbf{G}}$ is used to store the possible labels of each pixel)\\
\textbf{repeat}\\
\quad $\emph{\textbf{U}}$[1:$N$, 1:3] $\gets 0$, $\emph{\textbf{H}}$[1:$N$, 1:3] $\gets 0$, $\emph{\textbf{f}}$[1:$N$] $\gets 0$\\
\quad \textbf{for} $j = 1$ \textbf{to} $z$ \textbf{do}\\
\quad\quad \textbf{for} $i = 1$ \textbf{to} SeachRegion($c_j$) \textbf{do}\\
\quad\quad\quad $d = $ Distance($x_i, c_j$)\quad\quad //SLIC distance function\\
\quad\quad\quad \textbf{if} $f(i)<3$ \textbf{then}\\
\quad\quad\quad\quad $f(i)$++, $D(i, f(i)) = d$, $G(i, f(i)) = i$\\
\quad\quad\quad \textbf{else}\\
\quad\quad\quad\quad \textbf{if} $d<\max{D(i, :)}$ \textbf{then}\\
\quad\quad\quad\quad\quad $D(i, $Position($\max{D(i, :)}$)$) = d$, $G(i, $Position($\max{D(i, :)}$)$) = i$\\
\quad\quad\quad\quad \textbf{end if}\\
\quad\quad\quad \textbf{end if}\\
\quad\quad \textbf{end for}\\
\quad \textbf{end for}\\
\quad Update $\emph{\textbf{U}}$\\
\quad Use $\emph{\textbf{H}}$ to get new $\emph{\textbf{U}}$ ($\emph{\textbf{U'}}$)\\
\quad Use $\emph{\textbf{U'}}$ to update $\emph{\textbf{c}}$\\
\textbf{until} Convergence or reaching the max number of iterations\\
\STATE
$\emph{\textbf{L}}= G($Position$(\max{(\emph{\textbf{U'}})}))$\\
\STATE 
Enforce the connectivity of $\emph{\textbf{L}}$\\
\STATE 
$m_p=$ Count(Unique($\emph{\textbf{L}}$))\\
\end{algorithmic}
\end{algorithm}

\subsection{Onion Peeling Algorithm} 
\begin{figure*}
\centering 
\centerline{\includegraphics[width=\linewidth]{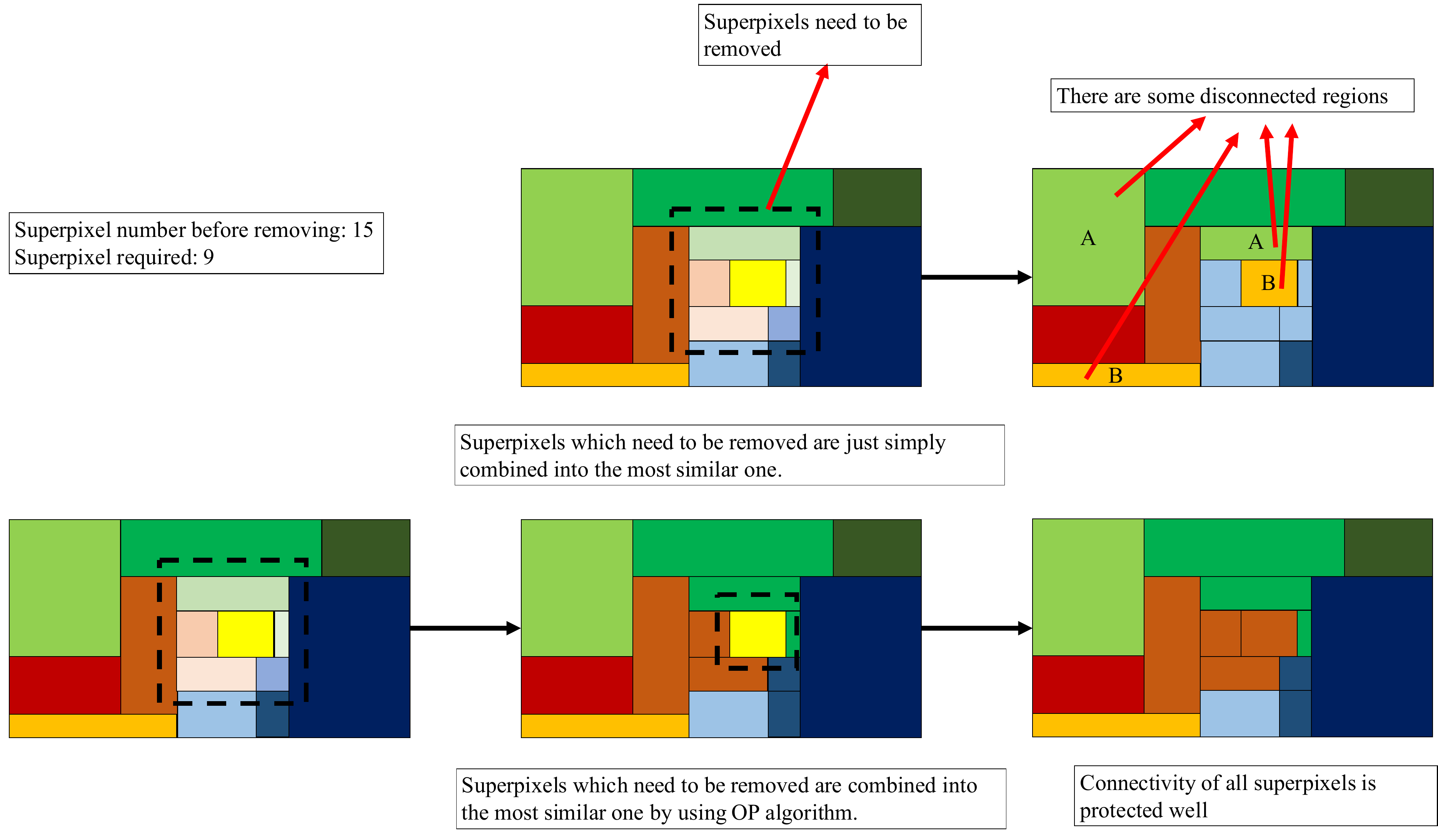}}
\caption{The combination and removing process of simple combination algorithm (Upper) and onion peeling (OP) algorithm (Bottom), where each grid represents a superpixel and each color represents a label. In simple combination algorithm, superpixels need to be removed are just combined into the most similar one which will cause some disconnected regions. To keep the connectivity, in OP algorithm when a superpixel needs to be removed if it is only surrounded by other superpixels that need to be removed, it will be processed after one of its neighbor processed. Because this process is quite like peeling onion from outside to inside, the algorithm is called the onion peeling algorithm.
}
\label{fignx}
\end{figure*}

Clustering-based superpixel methods often need a post-processing step to enforce the connectivity. However, this process leads these methods unable to generate precise number of superpixels and this defect limits the applications of these methods. In many cases, a precise number of superpixels is needed \cite{SPTPT}. In this paper, to solve this problem of clustering-based methods, we propose an algorithm called the onion peeling (OP) algorithm. When the number of generated superpixels is not the number as required, OP algorithm will rerun the clustering to generate more superpixels or combine some superpixels to reach the required number. The details of OP algorithm are shown in Fig. \ref{fignx} and Algorithm \ref{alg5}. In OP algorithm, to generate more superpixels, the algorithm needs to calculate the new cluster number which will be used in superpixel methods as follow,
\begin{equation}
\label{e9}
\mathcal{M} = \frac{m^2} {M}+mA,
\end{equation}
where, $M$ is the number of superpixels after enforcing connectivity, and $A$ is an amplification coefficient. In this paper, it is set to 0.2. 

The criteria to determine which superpixel will be removed is based on its size. Superpixels will be sorted by their sizes first. Then, superpixels will be removed by the ascending order according to their sizes until required number of superpixels obtained. The removed superpixel will be combined into the most similar one of its directly connecting neighbors. The criteria to determine the similarity between two directly neighboring superpixels is based on the L2-norm, which can be calculated as follow,
\begin{equation}
\label{e10}
d = \|c_i-c_j\|_2,
\end{equation}
where, $c_i$ is the centroid of superpixel $i$ and $c_j$ is the centroid of superpixel $j$.

\begin{algorithm}[htb]
\scriptsize
\caption{Onion peeling algorithm}
\label{alg5}
\begin{algorithmic}
\REQUIRE 
Required number of superpixels $z$, number of clusters $m$, number of superpixels after enforcing connectivity $M$, amplification coefficient $A$, label matrix $\emph{\textbf{L}}$\\
\ENSURE 
Label matrix $\emph{\textbf{L}}$, number of superpixels $M$\\
\STATE 
\textbf{if} $M \neq z$ \textbf{then}\\
\quad \textbf{if} $M < z$ \textbf{then}\\
\quad\quad Compute the new cluster number $\mathcal{M}$ and generate more superpixels\\
\quad \textbf{end if}\\
\quad $dif = M-z$\\
\quad Select some superpixels with the smallest area according to the $dif$\\
\quad Combine them into the nearest and most alike neighbor, and delete their labels\\
\textbf{end if}\\
\end{algorithmic}
\end{algorithm}

\begin{figure*}
\centering 
\centerline{\includegraphics[width=\linewidth]{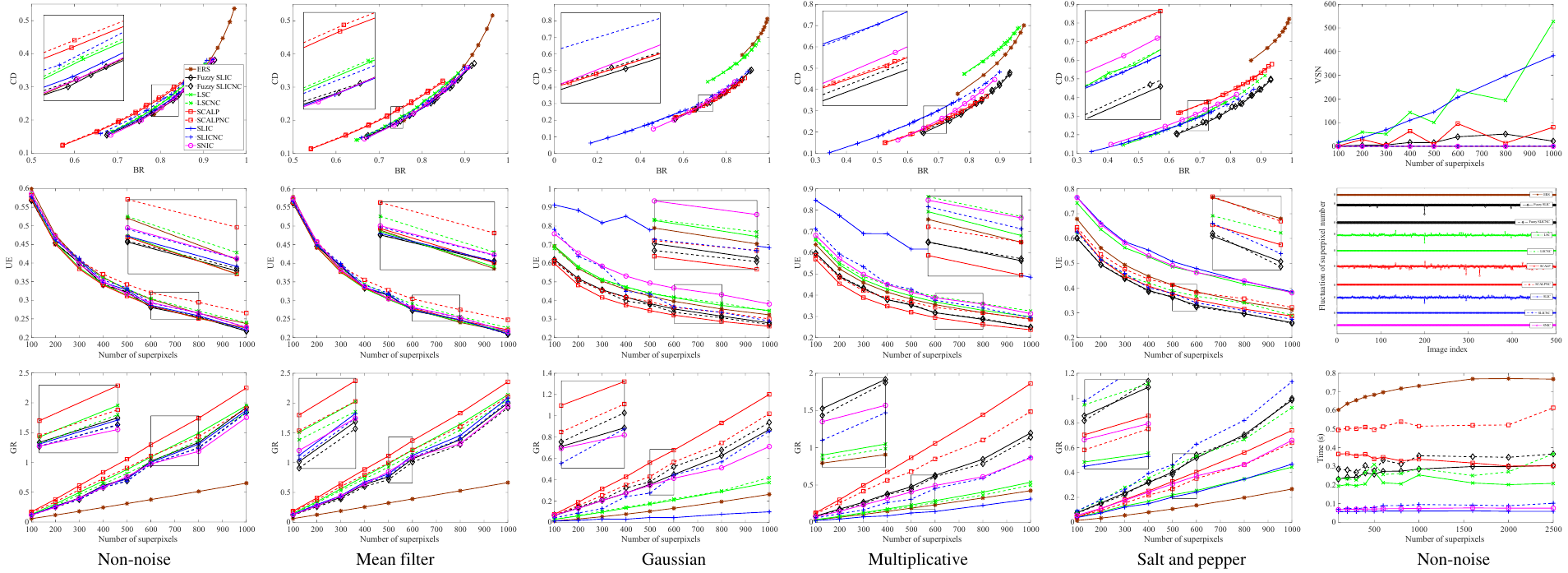}}
\caption{Overall performance comparison of all methods in BSD500 dataset. The noise level for Gaussian, multiplicative, and salt and pepper noise is 0.3. The average running time is obtained by each method generating different number of superpixels over 100 images of size $321\times481$.}
\label{fig2}
\end{figure*}

\subsection{Computational Complexity}
Fuzzy SLIC has to update $\emph{\textbf{U}}$ (fuzzy partition matrix), $\emph{\textbf{H}}$ (spatial information matrix), and $\emph{\textbf{U’}}$ (new fuzzy partition matrix) in its clustering method. The computational time of $\emph{\textbf{U}}$ and $\emph{\textbf{U’}}$ update is proportional to $N$, where $N$ is the number of pixels. The computational time of $\emph{\textbf{H}}$ update is proportional to $8\times N$ (8 neighbors of each pixel selected by a $3\times3$ window). The total computational time of Fuzzy SLIC is $O(11N) + O(z)$ which is comparable to SLIC ($O(N) + O(z)$), because $N \gg 10$. The computational complexity of OP algorithm is $O(z - M) + O(N)$, where $z$ is the expected superpixel number. In most cases, the complexity is $O(z - M) \ll O(N)$ since clustering-based methods are usually over-sampling (generated number $\geq$ the required) as Fig. 5 (the last diagram in the second row) shows.

\section{Experiments and Results}
\label{Exp}
\subsection{Experiment Settings}
To validate the proposed methods, we compared Fuzzy SLIC with ERS, LSC, SCALP, SLIC, and SNIC on BSD500 and Pascal VOC2007 benchmarks, and to ensure the fairness, we set the main parameters of them to the default. For the methods which have the regularity parameter, to make them comparable, we set parameter (Fuzzy SLIC 15, LSC and SCALP 0.3, SLIC and SNIC 20) to achieve similar UE in the noise-free environment. It should be noted that Fuzzy SLIC can achieve more robust results with a higher regularity parameter. A higher regularity parameter can produce a poorer performance in the noise-free environment. To maintain the consistency and fairness, we keep the regularity parameter the same because we also want to achieve the state-of-the-art performance in all situations including noise-free environment when our method is compared with existing ones. The performance of Fuzzy SLIC in the paper can be regarded as a baseline setting of Fuzzy SLIC. To test the robustness of all methods in this paper, we applied them and compared their performance under 5 noise conditions: non-noise, additional Gaussian noise with zero mean standard deviation (std) range $\left[0.1,0.2, 0.3\right]$, multiplicative Gaussian noise (mean 0 and std range $\left[0.1,0.2, 0.3\right]$), salt and pepper noise with density range $\left[0.1, 0.2, 0.3\right]$, and a simple mean filter, respectively. To test the OP algorithm, we also integrate it into Fuzzy SLIC and three clustering-based methods: LSC, SCALP, and SLIC. The integrated methods are called Fuzzy SLICNC, LSCNC, SCALPNC, and SLICNC respectively. We also tested the effect of different hyperparameters of Fuzzy SLIC on its performance. Experiments are run on a personal computer with Mac OSX 10.14.1, Intel Core i5 2.3 GHz 4 cores CPU, and 16 GB RAM.

\subsection{Benchmark Metrics}
The evaluation metrics we select are: global regularity (GR) \cite{GR}, contour density over boundary recall rate (CDBR) \cite{SCALP,GR,ASA} and under segmentation error (UE) \cite{Turbo,Vek}. We also use the variance of superpixel numbers (VSN) to evaluate the superpixel number control ability of each method. VSN calculate the variance of generated superpixel numbers in a set of same size images using the same $m$.

\paragraph{Boundary recall (BR)} BR is used to evaluate the boundary adherence of superpixel methods. It measures how many boundary pixels of ground truth can be matched by the boundary pixels of a superpixel. If a ground truth boundary pixel is within a distance threshold (usually 2 pixels) of a superpixel boundary pixel, it can be regarded as a hit \cite{Ren2003}. The percentage of ground truth boundary pixels matched by superpixel boundaries is the recall rate, and it is ranged from 0 to 1, the higher boundary recall rate means better boundary adherence. Boundary recall rate can be calculated using following formula,
\begin{equation}
BR = \frac{1}{N_g}(\sum_{i=1}^{Ng}{\rm logical}(\min_{y_j}\|x_i-y_j\|\leq2)),
\end{equation}
where, $N_g$ is the total number of ground truth boundary pixels, $y_j$ is the spatial position of one pixel in superpixel boundary pixel set, and $x_i$ is the spatial position of ground truth boundary pixel $i$.

\paragraph{Contour density (CD)}
Since BR metric only considers the true positive detection and ignore the density of produced superpixel contours, using this metric only will cause a situation where the shape of superpixel is irregular although it has a high BR value. To overcome this problem, contour density (CD) is proposed to penalize a large number of superpixel boundaries $\mathcal{B(S)}$ \cite{ASA}, the formula of CD is as follow,
\begin{equation}
\begin{split}
CD =  \frac{|\mathcal{B(S)}|}{N},
\end{split}
\end{equation}
where, $|\mathcal{B(S)}|$ is the number of boundaries of superpixel $\mathcal{S}$, and $N$ is the total number of pixels. We use CD over BR (CDBR) \cite{SCALP,GR,ASA} to evaluate the performance of superpixel methods.

\paragraph{Under segmentation error (UE)} UE is another boundary adherence evaluation metric. In an ideal superpixel method, every superpixel should only belong to single object. UE measures the percentage of superpixels which belong to multiple objects. So it is also called leaking rate. It can be calculated as follow,
\begin{equation}
\begin{split}
UE=\frac{1}{N}(\sum_{j=1}^{z}{\rm logical}(o_j\mid (o_j\bigcap g_a) \ne \emptyset)*\\ {\rm logical}(o_j\bigcap g_a \ne \emptyset)*|o_j|),
\end{split}
\end{equation}
where, $o_j$ denotes the pixel set of superpixel $j$, $|o_j|$ is the number of pixels in superpixel $j$, $z$ is the number of superpixels, $g_a$ denotes the pixel set of any ground truth object, and ${\rm logical}(o_j\mid (o_j\bigcap g_a)  \ne \emptyset)*{\rm logical}(o_j\bigcap g_a  \ne \emptyset) \equiv 1$ means that superpixel $j$ belongs to multiple objects.

\paragraph{Global regularity (GR)}
We use the global regularity (GR) \cite{GR} to evaluate the shape regularity of superpixels obtained by each superpixel method. The formula of GR is as follow,
\begin{equation}
GR(\mathcal{S}) = SRC(\mathcal{S})SMF(\mathcal{S}),
\end{equation}
where, $SRC(\mathcal{S})$ is shape regularity criteria of the superpixel $\mathcal{S}$, and $SMF(\mathcal{S})$ is the smooth matching factor of the superpixel $\mathcal{S}$. The detailed formulas of $SRC$ and $SMF$ can refer to \cite{GR}.

\subsection{Performance Analysis}
Fig. \ref{fig2}\footnote{The results for other noise levels of BSD500 and the results of Pascal VOC2007 are included in supplementary material. Our methods achieve similar results of other noise levels in BSD500. Although their performance in Pascal VOC2007 is not as good as in BSD500, their overall performance is still better than other methods. Larger and clearer version of visual and quantitative comparison results are available in supplementary material.} shows the CDBR, UE, GR, VSN and average running time curves obtained by all methods used in this paper. We can see that Fuzzy SLIC and Fuzzy SLICNC achieve the best CDBR against state-of-the-art methods in all noise cases. It can be seen that Fuzzy SLIC and Fuzzy SLICNC achieve the second best GR which is just inferior to SCALP and SCALPNC or SLICNC in Gaussian noise, multiplicative noise, and salt and pepper noise. Fuzzy SLIC and Fuzzy SLICNC achieve the best or second best UE in all noise cases. We also can see that all methods integrated with the OP algorithm can control the superpixel number well. In addition, the performance of the methods after integration are similar or a little inferior to the original ones. The average running time of Fuzzy SLIC is better than SCALP which is another robust superpixel method. The running time of the methods after integration except SCALPNC is a little higher than the original ones. Visual comparison results of all methods are shown in Figs. \ref{fig19}-\ref{fig194}. We can see that in the cases of noise-free and mean filter, the difference between Fuzzy SLIC, Fuzzy SLICNC, LSC, LSCNC, SCALP, SCALPNC, SLIC, SLICNC, and SNIC is not significant. SCALP and SCALPNC achieve the best performance in Gaussian noise environment while Fuzzy SLIC and Fuzzy SLICNC achieve the second best performance. In the multiplicative noise environment, SCALP, SCALPNC and Fuzzy SLIC achieve the best performance and Fuzzy SLICNC achieve the second best performance. The difference between the first rank and the second rank is not significant. In the salt and pepper noise environment, Fuzzy SLIC and Fuzzy SLICNC achieve the best performance. LSCNC and SLICNC achieve the second best performance. However, the performance of SCALP is not stable in this noise and is much inferior to Fuzzy SLIC. And we can see an interesting phenomenon that some superpixel methods like LSC and SLIC after using OP algorithm are more robust than original ones. It shows that OP algorithm can improve the robustness of some existing superpixel methods to some extent.

\subsection{Ablation Study of Different Searching Range of Dynamic Fuzzy Superpixels}
\begin{figure*}
\centering 
\centerline{\includegraphics[width=\linewidth]{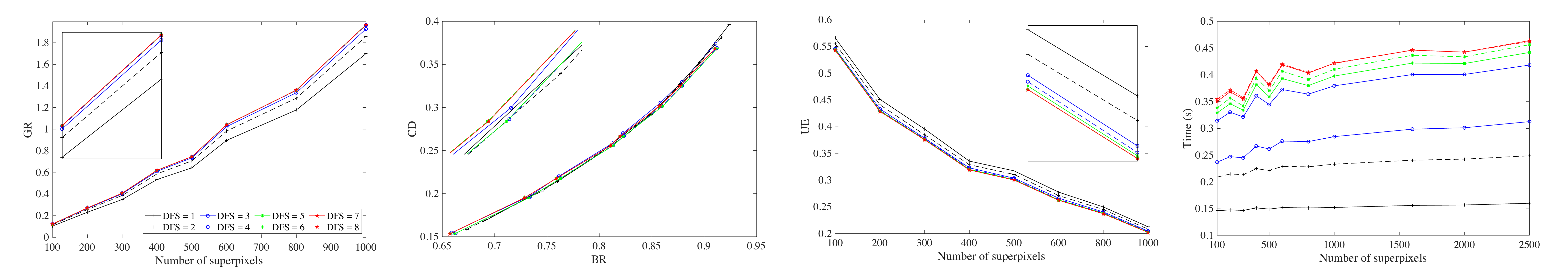}}
\caption{Comparison results of Fuzzy SLIC with different searching sizes of DFS on the BSD500 without noise.
}
\label{figx5}
\end{figure*}

To select the best searching size of dynamic fuzzy superpixels, we designed an ablation study. We shifted the maximum labels of DFS from 1 to 8 to see the effect of different searching sizes on the performance of Fuzzy SLIC. Fig. \ref{figx5} shows the comparison results of Fuzzy SLIC with different searching sizes of DFS on the BSD500 without noise. The CDBR of all comparison methods is almost the same, while there is a clear boundary in the curves of GR and UE: That is DFS = 3. We can see that Fuzzy SLIC with DFS = 1 and Fuzzy SLIC with DFS = 2 are significantly inferior to higher DFS searching sizes. There is no obvious difference between the performance of DFS = 3 and higher DFS searching sizes. What's more, the speed of DFS = 3 is much faster than higher DFS searching sizes. Hence, we use DFS = 3 in our final Fuzzy SLIC in main experiment.

\subsection{Ablation Study of Different $p$ and $q$ Allocations in Local Spatial Fuzzy C-means Clustering}
\begin{figure*}
\centering 
\centerline{\includegraphics[width=0.8\linewidth]{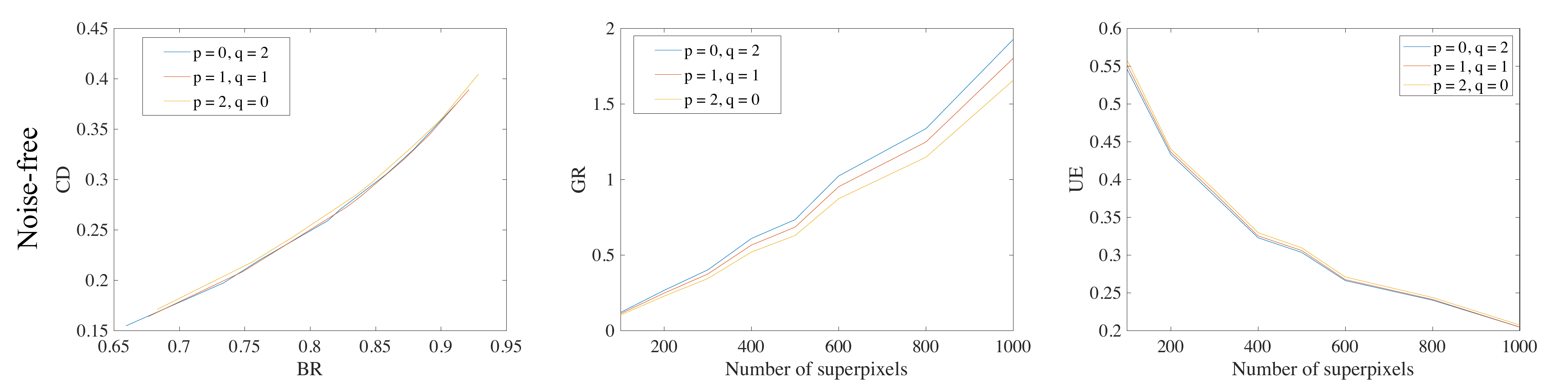}}
\caption{Comparison results with different $p$ and $q$ values in noise-free environment on the BSD500.
}
\label{figx1}
\end{figure*}

\begin{figure*}
\centering 
\centerline{\includegraphics[width=0.8\linewidth]{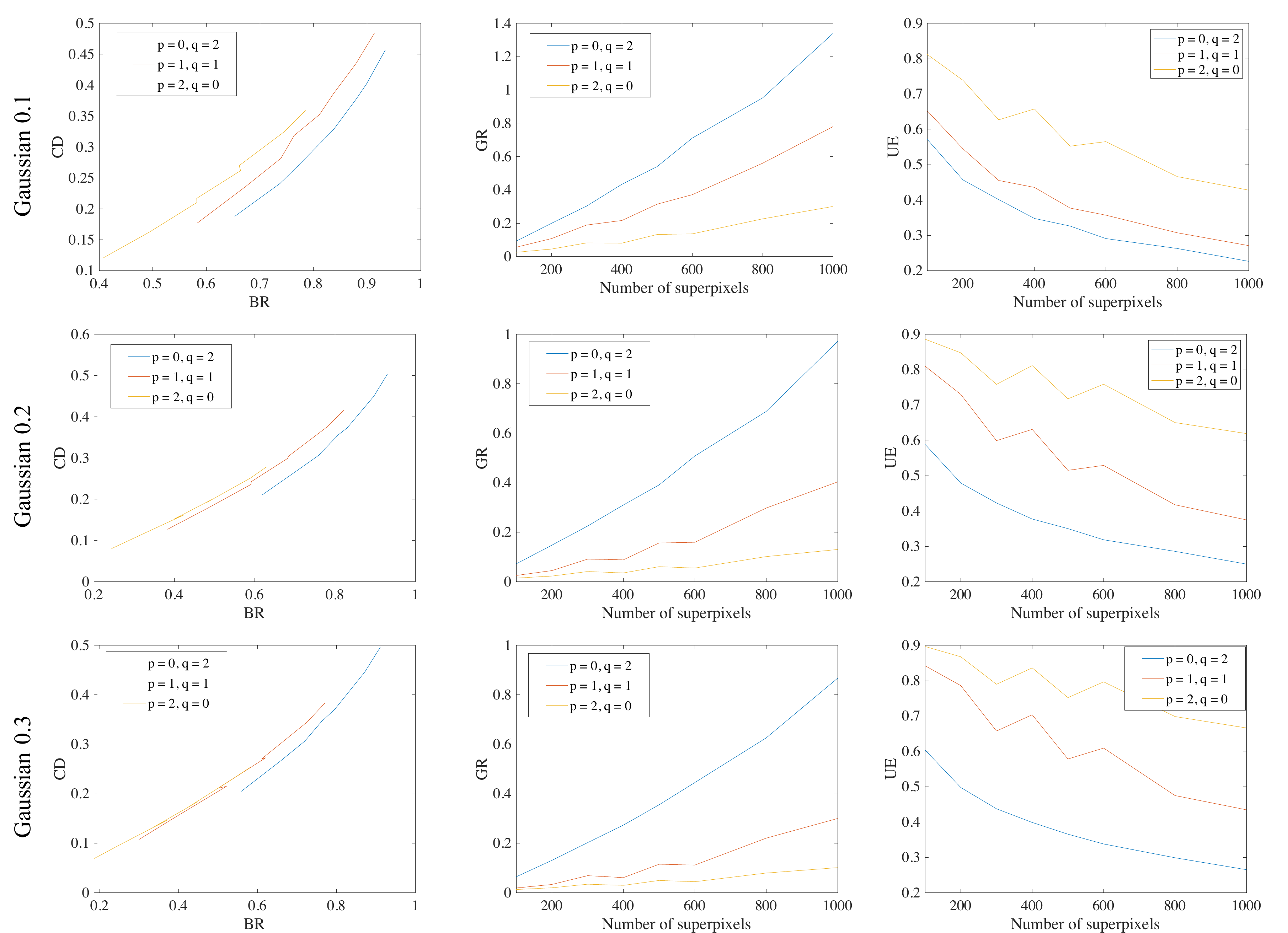}}
\caption{Comparison results with different $p$ and $q$ values with Gaussian noise on the BSD500.
}
\label{figx2}
\end{figure*}

\begin{figure*}
\centering 
\centerline{\includegraphics[width=0.8\linewidth]{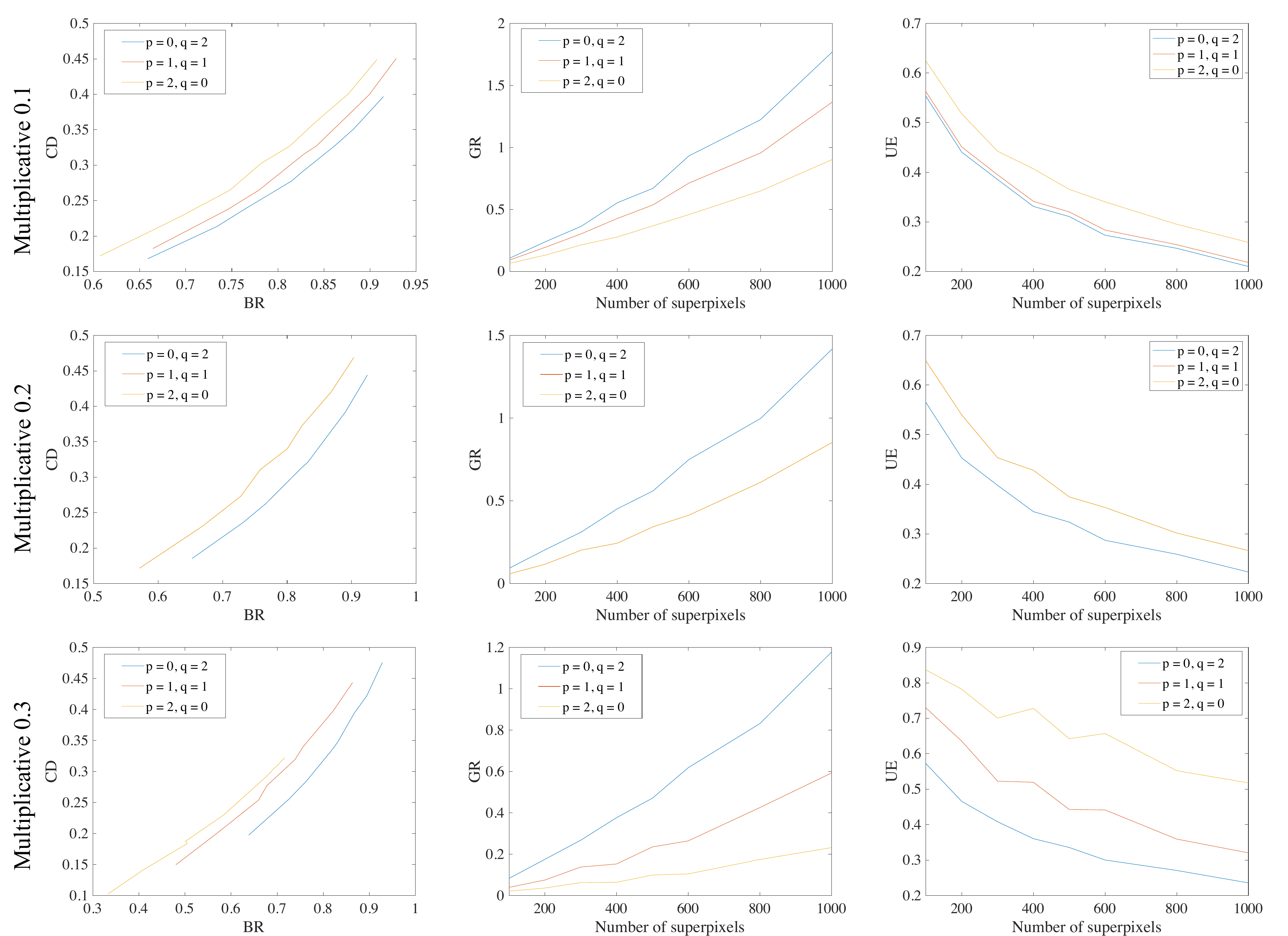}}
\caption{Comparison results with different $p$ and $q$ values with multiplicative noise on the BSD500.
}
\label{figx3}
\end{figure*}

\begin{figure*}
\centering 
\centerline{\includegraphics[width=0.8\linewidth]{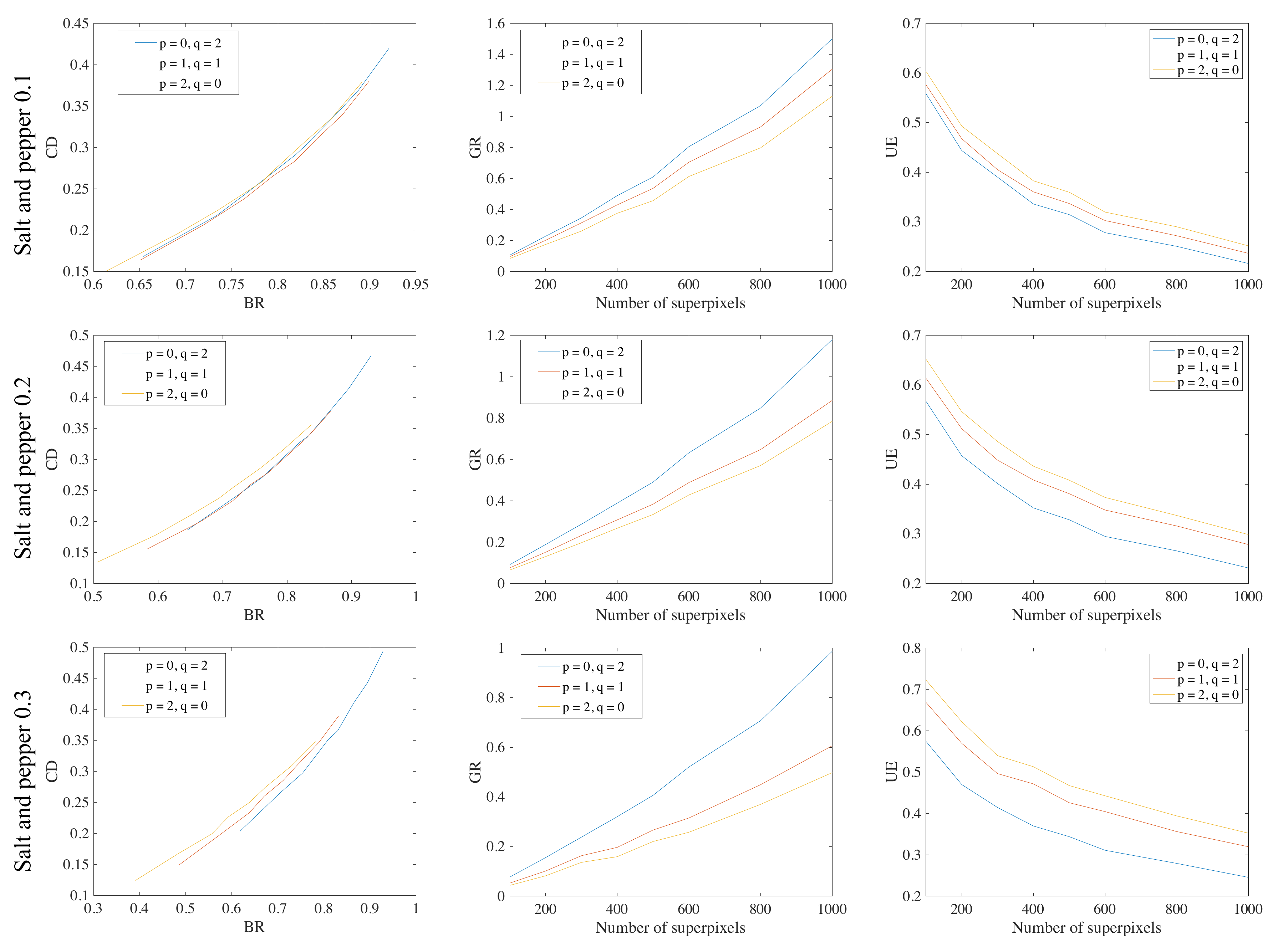}}
\caption{Comparison results with different $p$ and $q$ values with salt and pepper noise on the BSD500.
}
\label{figx4}
\end{figure*}

$p$ and $q$ in the local spatial fuzzy C-means clustering can affect the robustness and the performance of Fuzzy SLIC. To find the best allocation of $p$ and $q$, we designed this ablation study. As paper \cite{SFCM} did, we used 3 allocations of $p$ and $q$: (1) $p=0,q=2$, (2) $p=1,q=1$, (3) $p = 2, q=0$ in this study. Figs. \ref{figx1}-\ref{figx4} show the comparison results of different $p$ and $q$ in noise-free and noisy environment. In Fig. \ref{figx1}, we can see that in noise-free environment, all 3 allocations of $p$ and $q$ achieves similar performance on CDBR and UE while Fuzzy SLIC with $p=0,q=2$ achieves the best GR in noise-free environment. Fuzzy SLIC with $p=0,q=2$ outperforms other 2 allocations of $p$ and $q$ significantly in terms of CDBR, UE, and GR in noisy environment as Figs. \ref{figx2}-\ref{figx4} show.

\section{Conclusion}
This paper proposes a robust superpixel method using local spatial fuzzy C-means clustering and dynamic fuzzy superpixels. The proposed method, compared to state-of-the-art methods, has better performance in terms of boundary adherence, compactness, and computational complexity under a noise-free environment. More importantly, under a noisy environment, the proposed method outperforms state-of-the-art ones significantly. To solve the precise superpixel number control problem of most clustering-based superpixel methods, we propose the OP algorithm. In the validation experiments, the OP algorithm enables a clustering-based superpixel method to control the superpixel number fast and precisely.



\begin{IEEEbiography}[{\includegraphics[width=1in,height=1.25in,clip,keepaspectratio]{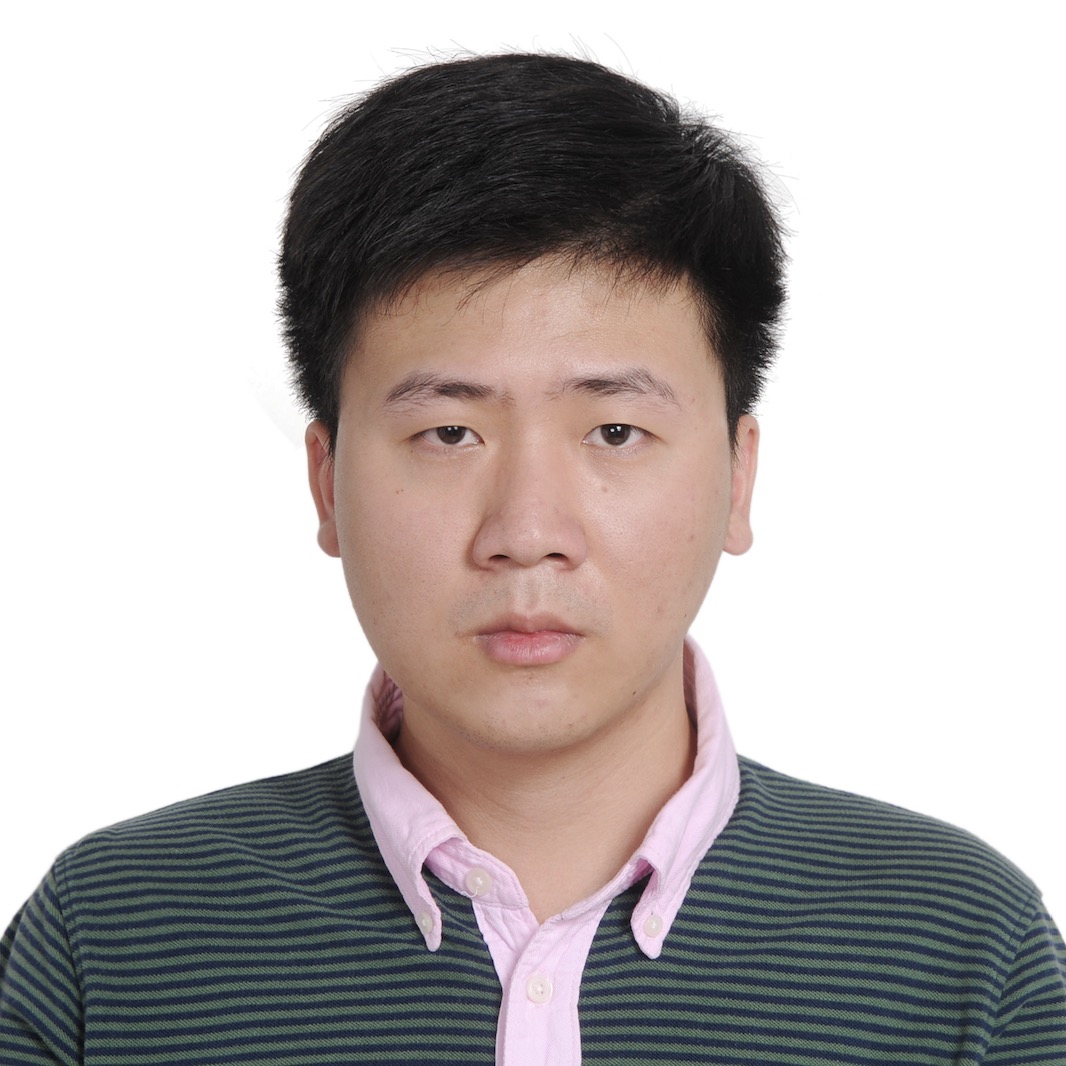}}]{Chong Wu} (S’19) was born in Zhejiang, China. He received the B.E. degree in automation from the School of Automation, China University of Geosciences, Wuhan, China, in 2018. He is currently pursuing the Ph.D. degree in electrical engineering with the Department of Electrical Engineering, City University of Hong Kong, Hong Kong. 

His current research interests include graph representation learning, image/video processing, and artificial intelligence. 
\end{IEEEbiography}

\begin{IEEEbiography}[{\includegraphics[width=1in,height=1.25in,clip,keepaspectratio]{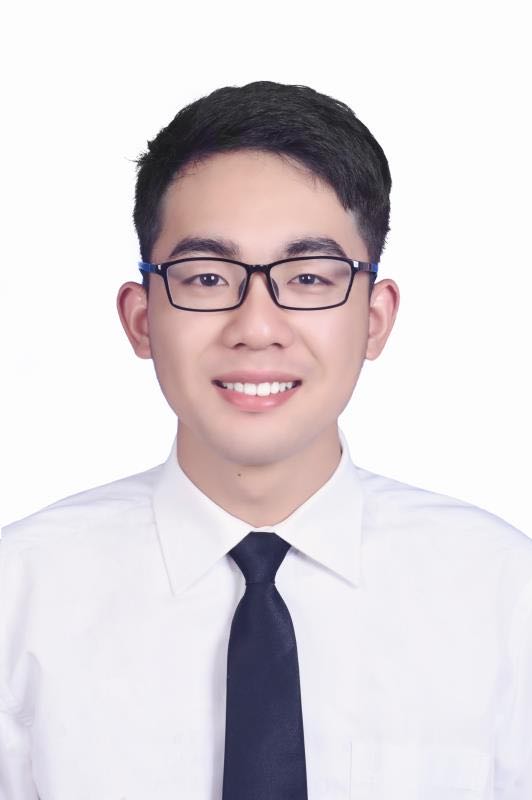}}]{Jiangbin Zheng} was born in Taizhou, Zhejiang, China, in 1996. He is currently pursuing the master's degree with Xiamen University, specializing in intelligent science and technology. 

His current research interests include cross-modal sign language translation, machine translation and graph neural network.
\end{IEEEbiography}

\begin{IEEEbiography}[{\includegraphics[width=1in,height=1.25in,clip,keepaspectratio]{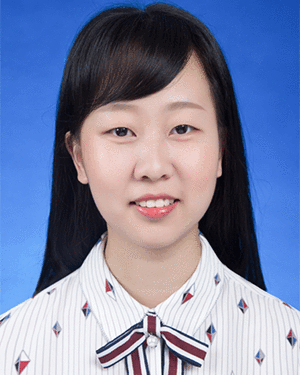}}]{Zhenan Feng} received the B.E. degree in automation in 2018 from the School of Automation, China University of Geosciences, Wuhan, China, where she is currently working toward the master's degree in control science and engineering.

Her current research interests include graph representation learning, image/video processing, motors and controls, design and optimization of electrical systems, and artificial intelligence.
\end{IEEEbiography}

\begin{IEEEbiography}[{\includegraphics[width=1in,height=1.25in,clip,keepaspectratio]{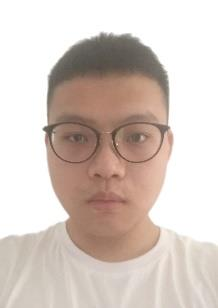}}]{Houwang Zhang} (S’19) received the B.E. degree in industrial design in 2018 from the School of Mechanical Engineering and Electronic Information, China University of Geosciences, Wuhan, China. Since then, he is currently working toward the master's degree in control science and engineering with the School of Automation, China University of Geosciences, Wuhan, China.

His current research interests include graph representation learning, image/video processing, and bioinformatics.
\end{IEEEbiography}

\begin{IEEEbiography}[{\includegraphics[width=1in,height=1.25in,clip,keepaspectratio]{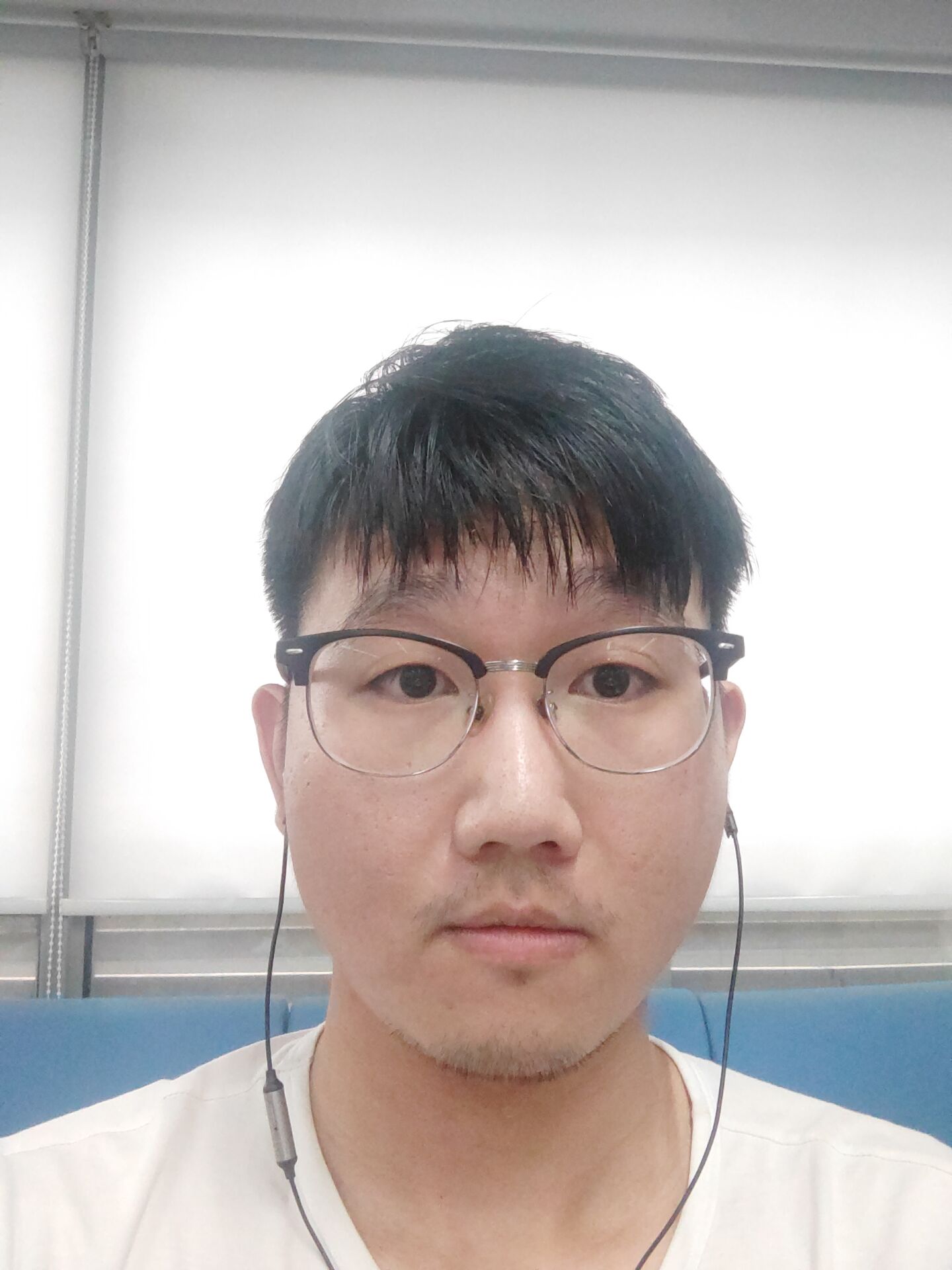}}]{Le Zhang} received the B.E. degree in software engineering from the Hangzhou Institute of Service Engineering, Hangzhou Normal University, Hangzhou, China, in 2018. He is currently pursuing the master’s degree with Tongji University. 

His current research interests include computer vision and big data.
\end{IEEEbiography}

\begin{IEEEbiography}[{\includegraphics[width=1in,height=1.25in,clip,keepaspectratio]{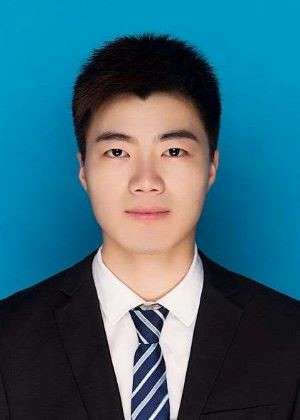}}]{Jiawang Cao} received the B.E. degree in automation from the School of Automation, China University of Geosciences, Wuhan, China, in 2019. He is currently pursuing the master’s degree with Fudan University. 

His current research interests include computer vision and deep learning.
\end{IEEEbiography}

\begin{IEEEbiography}[{\includegraphics[width=1in,height=1.25in,clip,keepaspectratio]{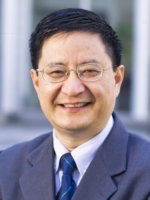}}]{Hong Yan} received the Ph.D. degree from Yale University. He was a Professor of imaging science with the University of Sydney. He is currently the Chair Professor of computer engineering with the City University of Hong Kong. He was elected an IAPR Fellow for contributions to document image analysis and an IEEE Fellow for contributions to image recognition techniques and applications. He received the 2016 Norbert Wiener Award from the IEEE SMC Society for contributions to image and biomolecular pattern recognition techniques.

His current research interests include bioinformatics, image processing, and pattern recognition.
\end{IEEEbiography}

\end{document}